\documentclass[aps,prb,twocolumn,float,a4paper,amsmath,amssymb]{revtex4}

\usepackage[dvips]{color}
\definecolor{g-blue}{rgb}{0.83,0.95,1}
\definecolor{g-yellow}{rgb}{1,1,0.7}
\definecolor{g-green}{rgb}{0.9,1,0.9}
\definecolor{green}{rgb}{0,0.6,0}
\definecolor{cyan}{rgb}{0,0.7,0.7}
\definecolor{black}{rgb}{0,0,0}
\definecolor{grey}{rgb}{0.4 ,0.4 ,0.4 }

 \usepackage{graphicx}
 \usepackage{amsmath,bm,epsfig}
\usepackage{verbatim}
\def \ed {\end{document}}
\def\Fbox#1{\vskip1ex\hbox to 8.5cm{\hfil\fboxsep0.3cm\fbox{%
  \parbox{8.0cm}{#1}}\hfil}\vskip1ex\noindent}  

\def\be{\begin{equation}}\def\ee{\end{equation}}
\def\bea{\begin{eqnarray}}\def\eea{\end{eqnarray}}
\def\bse{\begin{subequations}}\def\ese{\end{subequations}}
\newcommand{\BE}[1]{\begin{equation}\label{#1}}
\newcommand{\BEA}[1]{\begin{eqnarray}\label{#1}}
\newcommand{\BSE}[1]{\begin{subequations}\label{#1}}

\let \= \equiv \let\*\cdot \let\~\widetilde \let\^\widehat \let\-\overline

  \def\1{\bm1} 

\def\<{\left\langle}    \def\>{\right\rangle}
\def\({\left(}          \def\){\right)}
 \def \[ {\left [} \def \] {\right ]}

\newcommand{\B}[1]{{\bm{#1}}}

\renewcommand{\sb}[1]{_{\text {#1}}}  
\def\Sb#1{_{\scriptscriptstyle\rm{#1}}}

\begin{document}

\title{Energy Spectra of  Superfluid Turbulence in $^3$He}
\author{Laurent Bou\'e, Victor L'vov, Anna Pomyalov, Itamar Procaccia}
\affiliation{Department of Chemical Physics, The Weizmann Institute of Science, Rehovot 76100, Israel}

\begin{abstract}

In superfluid   $^3$He  turbulence is carried predominantly by the superfluid component. To explore the statistical properties of this quantum turbulence and its differences from the classical counterpart we adopt the time-honored approach of shell models. Using this approach we provide  numerical simulations of a Sabra-shell model that allows us to uncover the nature of the energy spectrum in the relevant hydrodynamic regimes.  These results are in qualitative agreement with analytical expressions for the superfluid turbulent energy spectra that were found using a differential approximation for the energy flux.

\end{abstract}

\date{\today}

\maketitle

\noindent
{\bf Introduction:} Superfluids are known to display many spectacular properties, such as the fountain effect and rolling films, which are unique to quantum hydrodynamics.  Ever since the discovery of superfluid it was conjectured that their
hydrodynamic can be described as two inter-penetrating fluids with two different velocities $v_s$ (superfluid component) and $v_n$ (normal component) \cite{balibarHist}.  The relative density of each component, ($\rho_s$ and $\rho_n$ respectively), depends on the temperature; this so-called ``two-fluid'' model was later developed very usefully  by Landau \cite{balibarHist}.  In this model the normal component is described by the Navier-Stokes equations and the superfluid part is modeled by the Euler equation.  Accordingly the inviscid super-component is characterized formally by an infinite Reynolds number: it is very easy for superfluids to become turbulent.  Quantum mechanics plays an imprtant role: unlike classical fluids where the circulation around the vortices is a statistical and dynamical variable, the properties of quantum vortices are tightly constrained by the laws of quantum mechanics. The circulation around a quantum vortex is quantized to integer values of  a fundamental unit called the quantum of circulation ~$\kappa = 2 \pi \hbar / m$, where~$m$ is the mass of a superfluid particle.  Because of this one can identify a new typical length-scale
in quantum turbulence, refereed to herewith as~$\ell$, which corresponds to the mean distance between quantum vortices.

At finite temperatures $T\neq0$ the normal and superfluid components are coupled to each other via thermal excitations and the resulting dissipative force is called the mutual friction.  This description is in qualitative variance with ordinary fluids and accordingly the question of how different classical and quantum turbulence are has emerged as a hot topic in the fluid mechanics community\cite{review1,review2}.  An important step in establishing a theory of quantum turbulence consists in developing a framework that describes the energy spectra of both the normal~$E_n(k)$ and the superfluid~$E_s(k)$ components.  The work to achieve this goal is still developing and there is a possibility that progress here  might also shed some light on some long standing classical turbulence problems.

Because of the high kinematic viscosity of~$^3$He (comparable to that of olive oil)\cite{ExpOlive}, the normal component can usually be considered to be at rest with respect to the superfluid component with~$v_n \ll v_s$.  In such circumstances we propose that the overall behavior of quantum turbulence is largely dominated by the superfluid velocity $\B v\equiv \B v_s$ component only.  Therefore, the description of the fluid motion can be reduced to the
coarse-grained hydrodynamic equation for the superfluid velocity $\B v(\B r, t)$, considered on length scales of the order of $l$ which are larger than intervortex distance~$\ell$:
\begin{subequations}\label{1-F}
\begin{equation}\label{Euler}
\frac{\partial {\bm v}}{\partial t} + \left({\bm v} \cdot \nabla \right) {\bm v} + \nabla \mu = - \B D  + \B f \ .
\end{equation}
Here $\mu$ is the chemical potential, $\B f(\B r, t)$ is a random force which is introduced in order to mimic the energy injection that sustains turbulence; this force acts at the outer scale $l_0\gg l\gg\ell$. The dissipative term $\B D$ represents the effective mutual friction which is the last remnant of the underlying but vanishingly small normal component.     Throughout the paper it should be kept in mind that the largest relevant wavevector is~$k_{\mbox{\tiny max}}= 1/\ell$; in other words, we restrict ourseleves to the pure hydrodynamic regime of superfluid turbulence.  In this case, the mutual friction can be approximated \cite{LNV} as
$\bm D \approx \alpha \, {\bm \omega} \times \left( {\bm \omega \times {\bm v}}  \right) / |{\bm \omega}|$.
Here  $\alpha$ is the temperature dependent dimensionless mutual friction parameter and~${\bm \omega} = \nabla \times {\bm v}$ is the superfluid vorticity.  This expression can be further simplified by averaging out the vectorial structure and recognizing that vorticity in hydrodynamic turbulence is dominated by the smallest possible eddies with~$k \sim 1/\ell$.  Because of their small size, these eddies also have very small turnover times.  On the other hand, the superfluid velocity~${\bm v}$ is dominated by the largest eddies who, in turn, have a very long turnover time.  This means that~${\B \omega}$ and~$\B v$ can be considered as almost completely uncorrelated.  In this case, we can replace~$\B \omega$ by its mean square value and the dissipation term can be approximated by
\begin{equation}\label{m-fric}
\B D\approx  \Gamma \B v\,,
\end{equation}\end{subequations} \vskip -0.8cm
\begin{equation}\label{self-con}
 \quad \Gamma\equiv  \alpha \, \omega\Sb T\,, \quad \omega^2\Sb T
\equiv  \<|\bm \omega|^2\>\approx 2 \int_{k_0}^{1/\ell}k^2 E(k)\, dk\ ,
\end{equation}
where $E(k)$ is the so-called 1-dimensional energy spectrum, normalized such that the total energy density per unite mass is $E=\int E(k) dk$.

Our goal below is to analyze the \emph{stationary} energy spectra of superfluid turbulence, $E(k)$,  at scales larger than the intervortex distance. This allows us to simplify further the problem by considering  $\Gamma$ in Eq.~\eqref{m-fric} as  a  time independent external parameter. Then, after finding $E(k)$ for a prescribed $\Gamma$ we can get the $\alpha$ corresponding to that $\Gamma$ using Eq.~\eqref{self-con}.  We will see that even after all these simplification the
spectra found below are highly non-trivial, justifying this study on its own right.

\begin{figure*}
\begin{center}
\includegraphics[width=0.345 \linewidth]{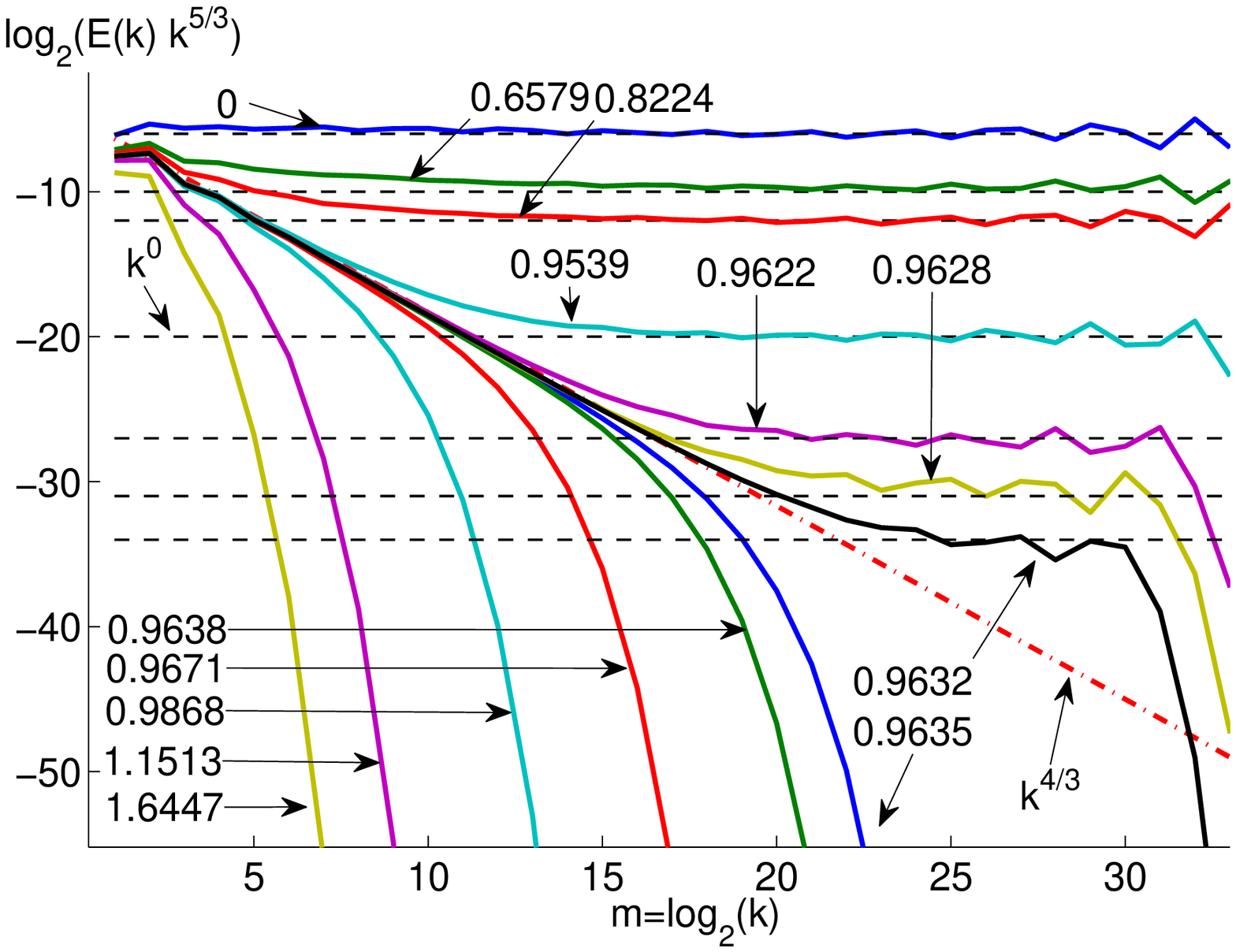}
\includegraphics[width=0.315 \linewidth]{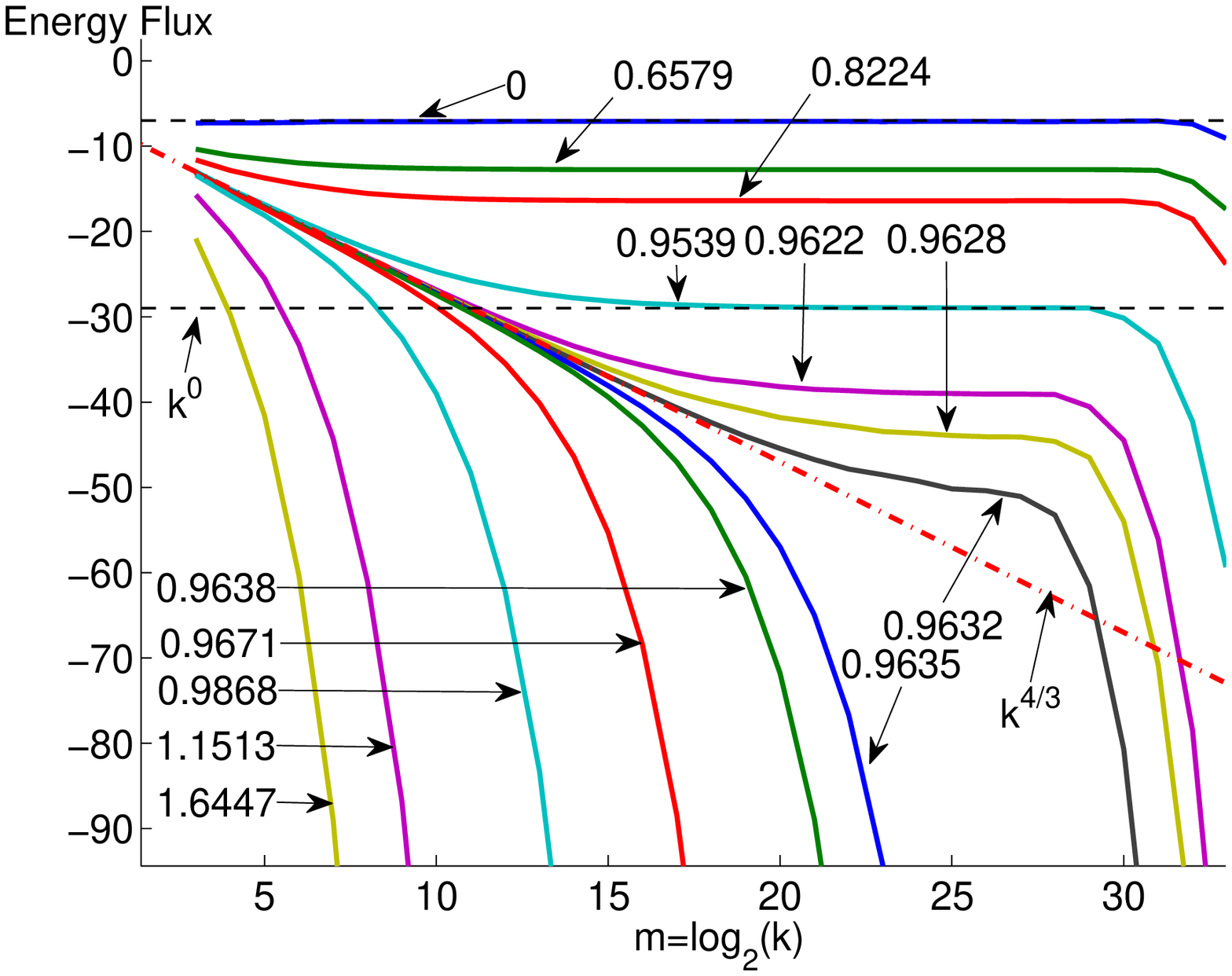}
\includegraphics[width=0.32 \linewidth]{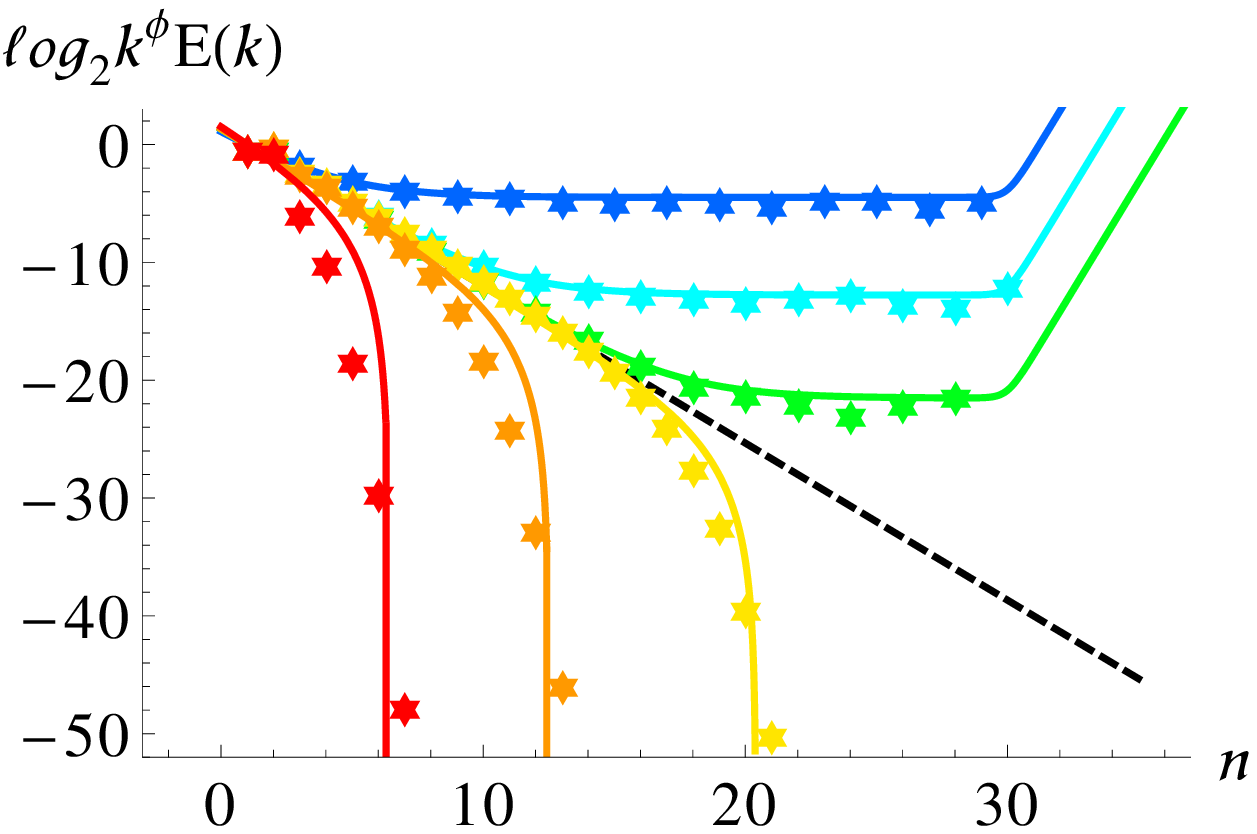}
\caption{\label{f:1} \emph{Left/Middle panel}: Numerical results for the energy spectrum (left) and the energy flux (middle) in the Sabra-shell model for different values of~$\gamma$. \emph{Right panel}: Analytical trial functions~$E\sb{sp}(k)$, Eq.(\ref{full-sp}), and~$E\sb{sb}(k)$, Eq.(\ref{solsB}),  in the differential energy-flux model. The stars show a comparison with data from the Sabra-shell model at similar values of the ratio~$\gamma/\gamma\sb{cr}$.  The bottleneck effect~$E\sb{sp}(k) \sim k^2$ at the highest wavevectors is removed from the Sabra-shell model by proper choice of parameters to simplify the  simulations. }
\end{center}
\end{figure*}

In Sec.~I we examine Eqs.~\eqref{1-F} using a shell model of the type that was widely used in studies of classical turbulence. Of course, the shell model is an uncontrolled approximation of Eqs.~\eqref{1-F}. Nevertheless our experience with
 classical turbulence is that they provide strong indications to the correct spectra associated with the full equations. This Section   presents numerical simulations that resolve the energy spectra $E\Sb{SM}(k)$ for different values of the mutual friction extending over an unprecedented range of wave vectors; see the left panel of Fig.~\ref{f:1}.
 We demonstrate that energy  spectra depend crucially on the value of dimensionless mutual-friction frequency
\begin{equation}\label{gamma}
\gamma\equiv \Gamma \big / \varepsilon_0^{1/3} k_0^{2/3}\,,
\end{equation}
which is normalized by the energy influx $\varepsilon_0$ at the outer wave vector $k_0$. There is a critical value $\gamma\sb{cr}\approx 1$ at which $E(k)=E\sb{cr} \propto k^{-3}$.  For $\gamma<\gamma\sb{cr}$ one has supercritical spectra $E\sb{sp}(k)>E\sb{cr}$ (at the same values of $\varepsilon_0$ and $k_0$) with Kolomogorov-41 (K41) tail $E\sb{sp}(k)\simeq \epsilon^{2/3}k^{-5/3}$ and smaller energy flux $\varepsilon < \varepsilon_0$.  For $\gamma>\gamma\sb{cr}$ one has subcritical spectra $E\sb{sb}(k)<E\sb{cr}$ that  terminate at final  $k$.

To increase our trust in this complex behavior we reanalyze (in Sec.~II) the same basic Eqs.~\eqref{1-F} analytically, using approximate differential closures for the energy flux over scales. These studies culminate with analytical solutions Eqs.~\eqref{full-sp} and \eqref{solsB} for the super- and subcritical energy spectra respectively.  The closure techniques are not exact either, but we will see that they agree very well with the results of the shell model. The analytical spectra for $E\sb{sp}(k)$ and $E\sb{sb}(k)$ are shown in the middle panel of Fig.~\ref{f:1} where the stars correspond to the numerical data from Sabra-shell model with similar values of the ratio~$\gamma/\gamma\sb{cr}$.  One sees very good qualitative agreement between numerical and analytical results, despite the fact that both approaches are (uncontrolled) approximations of a very different nature. We propose that our results for the energy spectra can be considered as valid predictions on a semi-quantitative level.  A detailed analysis of our results is presented in the final~Sec.~III.

\vskip 0.4cm

\noindent {\bf I. Superfluid energy spectra in shell models}
\vskip 0.2cm

Here we investigate the superfluid energy spectra numerically in the shell-model approximation. Competing methods, such as vortex filament simulations or the solution of the non-linear Schr\"odinger equation are extremely demanding computationally, having therefore a very short span of scales. Shell models offer an attractive alternative, as recently suggested by Wacks and Barenghi \cite{WB2011} who proposed a two-fluid shell model for superfluid turbulence using a GOY shell model\cite{S1,S2} with additional coupling terms.   While only 18 shells were used in Ref.~[\onlinecite{WB2011}], we have been able to expand the range of scales quite significantly by increasing the number of shells to~36.  This improvement allows us to probe an inertial interval over scales which extends to more than five decades, thereby making it possible to resolve subregions with different scaling behavior. Also, instead of using a GOY model, which suffers from the artifact of ``period-three" oscillations in the energy spectrum, we used the Sabra shell model~\cite{Sabra} in which this non-physical phenomenon is removed and the scaling behavior become more transparent. These improvements allow us to get essentially new information about the statistical behavior of superfluid turbulence.

\vskip 0.3cm

\noindent {\bf \small \emph{A. ``General" shell models of hydrodynamic turbulence}}
\vskip 0.2cm

Shell models of turbulence~\cite{S1,S2,S3,S4,S5} are simplified caricatures of the equations of fluid mechanics~\eqref{Euler} in wave-vector representation: they mimic the statistical behavior of $\B k$-Fourier components of the turbulent velocity field $\B u(\B k, t)$ in the entire \emph{shell} of wave-vectors $k_m <  k_{m+1}$ by only one complex \emph{shell velocity} $u_m$.  The integer index $m$ is referred to as the shell index and the \emph{shell wave numbers} are chosen as a geometric progression $ k_m =k_0 \lambda ^m$,  with $\lambda>1$ being the ''shell spacing'' parameter; often  one chooses $\lambda=2$.  The equation of motion of the fluid [e.g. Fourier transform of Eq.~\eqref{Euler}] is reduced to a dynamical equation of motion for $u_m(t)$:

\begin{subequations}\label{S}\begin{eqnarray}
 \Big[\frac{d}{dt}+\Gamma \Big]u_m =\mbox{NL}_m\{u_j\}+f_m\ .
 \end{eqnarray}
The model contains a time-independent mutual friction parameter $\Gamma$ and random force terms $f_m $ whose contribution as the energy influx at large scales is usually limited to the first few shells only. The nonlinear term $\mbox{NL}_m\{u_j\}$, which originates from the $(\B v \cdot \B \nabla)\B v $ in Eq.~\eqref{Euler}, should be i) proportional to $k_m$ (because of $\B \nabla$), ii) quadratic in the set of velocities $\{u_j\}$ and iii) should conserve the total kinetic energy. In order to implement Richardson-Kolmogorov idea of a step-by-step cascade energy transfer over scales, the set of the shell indexes $j$ in $\mbox{NL}_m\{u_j\}$ should be localized around~$m$. As as rule, one takes $j=m, \   m\pm 1,\  m \pm 2$.


\vskip 0.3cm

\noindent  {\bf \small \emph{B.  Sabra-shell  model of turbulence} }
\vskip 0.2cm

It is known that the original \emph{GOY-shell model}~\cite{S1,S2} suffers from non-physical ``period-three" oscillations of the energy spectrum that originate from artificial phase correlations in subsequent triads of the shell velocities.   These oscillations were eliminated in the  \emph{Sabra-shell model}~~\cite{Sabra} by a proper choice of the nonlinear term:
\begin{eqnarray}\label{Sb}
\mbox{NL}_m\{u_j\}&=&i \big(a~ k_{m+1}u_{m+2}u^*_{m+1}\\\nonumber
&+&b k_m  u_{m+1} u^*_{m-1}-c k_{m-1}u_{m-1}u_{m-2}\big)\,,
 \end{eqnarray}\end{subequations}
where~$^*$ stands for complex conjugation and where the coefficients $a,b$ and $c$ are real. Conservation of energy in the forceless inviscid limit requires that $a+b+c=0$. It is known \cite{S1,S2,Sabra,S3,S4,S5}
 that the statistical properties of the velocities in  shell models mimic those of the three-dimensional hydrodynamic turbulence for $a=1,\ b=c=-0.5$. In the forceless inviscid limit the model has two quadratic invariants: the energy $E$ and   $H$, often associated with helicity:
 \begin{subequations}\label{I}
  \begin{eqnarray}\label{Ia}
 E=\sum_m E_m, \    H=  2\sum_m \Big(\frac a c\Big )^ n  E_m,\quad  E_m\equiv \frac {|u_m |^2} 2 .~~~~
  \end{eqnarray}
The 1D turbulent energy spectrum $E(k)$ can be connected to $E_m$ by the relation $E(k_m)=\langle E_m \rangle /k_m$.
  The energy flux over scales $\varepsilon_m$ in the Sabra model is:
\begin{equation}\label{eps}
  \varepsilon(k_m)= 2 {\rm Im}[ a k_{m+1} S_{m+1} -c k_m S_m]\,,
  \end{equation}
  \end{subequations}
where $ S_ m= \langle u_{m-1}u_m  u^*_{m+1} \rangle $ is the third-order correlation function and  Im$[\dots]$ denotes imaginary part of $[\dots]$.
\vskip 0.3cm

 \noindent  {\bf \small  C. \emph{ Numerical realization of the Sabra model} }
  \vskip 0.2cm

The set of $N=36$ coupled Eqs.~\eqref{S} was solved using 4th order ETD-Runge-Kutta method \cite{CoxMatthews02}
 for stiff systems.  In order to mimic the energy sink at very high wavenumbers (e.g. due to Kelvin waves), a hyperviscosity term of the form $\nu k^4 |u_m|^2$ with $\nu=10^{-32}$ was added to the equation.  Energy pumping into the system was carried out by the ``helicity-free'' forcing~$f_m$ \cite{Sabra}.  We used random forcing ($\delta$-correlated in time ), with a finite amplitude  and a random phase in the first two shells, such that the forcing amplitudes are related to each other as $f_2=\sqrt{-(c/a)} f_1$  and $f_1=0.01$.  Such a forcing minimizes the input of helicity  into the system.  The mutual friction coefficient $\gamma$  was chosen in the range $0\le \gamma < 1.65$ to expose the different solutions exhibited by the system. The use of hyperviscosity allowed to maximize the extent of the inertial interval. However the sharp onset of the viscous term lead to artificial oscillations in the energy spectra in a few shells around the dissipative cutoff (see Fig \ref{f:1},~left). These oscillations are an artifact of the numerical procedure and not the result of the underlying dynamics.

 Results of the numerical simulations  of Sabra Eqs.~\eqref{S}, shown in Figs.~\ref{f:1} and \ref{f:2}, will be discussed below in Sec.~III.

\vskip 0.4cm

 \noindent {\bf II. A differential  model for the energy spectra}
 \vskip 0.2cm
 \noindent  {\bf \small \emph{A. Energy balance equation with differential closure } }
  \vskip 0.2cm

  \noindent The energy spectrum of pure hydrodynamic turbulence can also be studied analytically by investigating the energy balance equation
\begin{equation}
\frac{d \varepsilon(k)}{d k}+\Gamma E(k)=0\,,
\label{contHydro}
\end{equation}
 which is presented here in the stationary case (i.e.~$ \partial E(k)/ \partial t \equiv 0$).  The energy density $E(k)$ is related to the total energy~$E$ by:
$ E=\int E(k) d k$.
Equation~\eqref{contHydro} directly  follows from Eqs.~\eqref{1-F} with some expression for the energy flux over scales,  $\varepsilon(k)$, in terms of the third order correlation function of $\B u_{\B k}$ exactly as in the case  of classical turbulence.  In order to proceed further one can borrow a closure procedure from classical turbulence that expresses $\varepsilon(k)$ in terms of the energy spectrum, $E(k)$.  Even though this step is widely used, it is worth remembering that it is uncontrolled.  The simplest \emph{algebraic} closure relation  suggested by Kovasznay~\cite{Kov-47}
\begin{subequations}\label{closure}
\begin{equation}\label{closureA}
\varepsilon(k)\simeq  \frac 58 k^{5/2} E^{3/2}\,,
\end{equation}
just follows from K41 dimensional reasoning. The prefactor~$\frac 58$ is chosen to simplify the appearance of some of the equations below.   Notice that this simple algebraic approximation~\eqref{closureA} fails to include an important  effect: the bottleneck energy accumulation near the inter-vortex scale $\ell$.  We therefore use the Leigh \emph{differential} closure~\cite{Leith67} for the energy flux in the form suggested by Nazarenko~\cite{Nazar-Leith}:
\begin{equation}\label{LN}
\varepsilon (k) = - \frac{1}{8} \sqrt{k^{11} E } \frac{d }{d k} \Big[ \frac{E (k)}{k^2} \Big]\ .
\end{equation}
\end{subequations}
Dimensionally, this closure relation coincides with Eq.~\eqref{closureA}, but replaces
 $E(k)/k^3$ with $d [E(k)/k^2]/ d\, k$.  This allows one to account for the existence of thermodynamic equilibrium when $\varepsilon(k)=0$ and $E(k)\propto k^2$. The prefactor~$-\frac 18$ is chosen to get a frictionless (i.e. with $\Gamma=0$)  K41 solution,
 \begin{subequations}\label{blocks}
 \begin{eqnarray}\label{K41}
E\Sb{K41}(\varepsilon|k) \equiv C\Sb{K}  \varepsilon^{2/3} k^{-5/3} ,  \   C\Sb{K}= \Big(\frac{24 }{ 11 }\Big)^{2/3}\approx 1.68\,,~~~
\end{eqnarray}
with $C\Sb K$ numerically close to its experimental value.

\begin{figure*}
\begin{center}
\includegraphics[width=0.325 \linewidth]{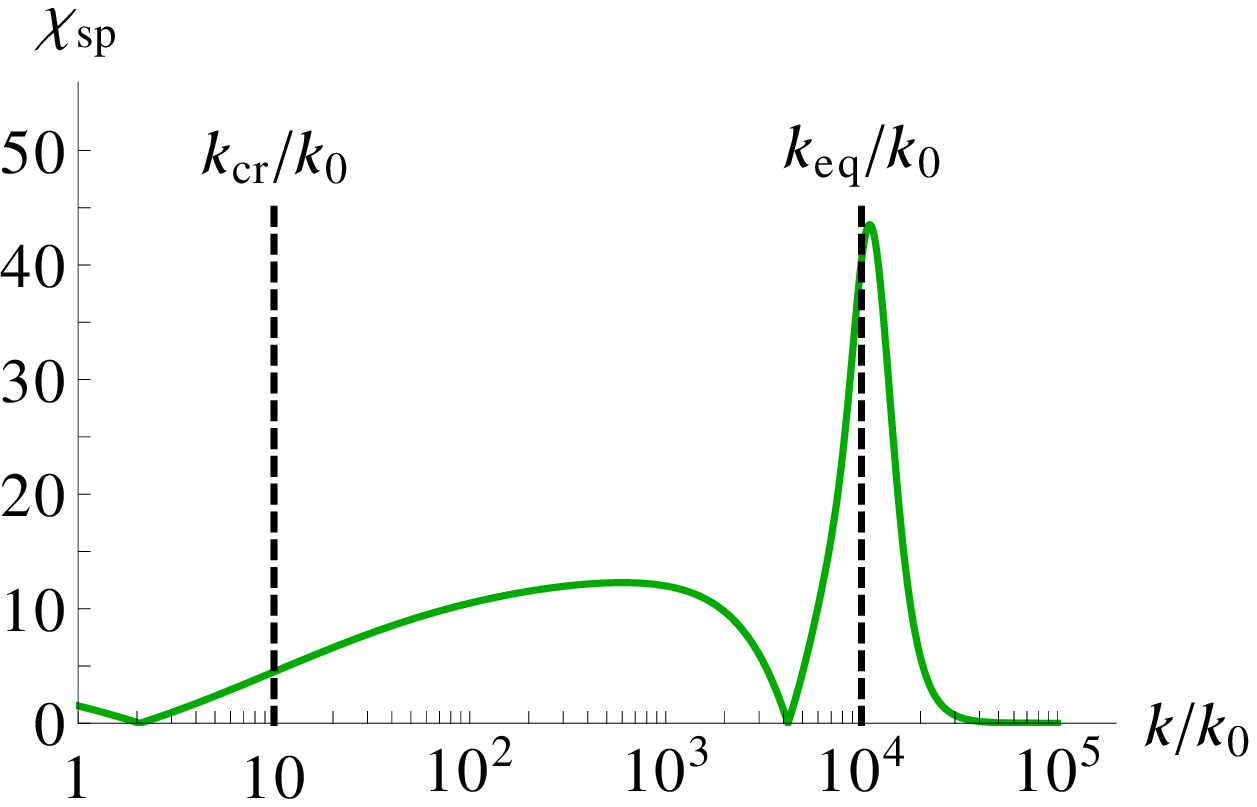}
\includegraphics[width=0.325 \linewidth]{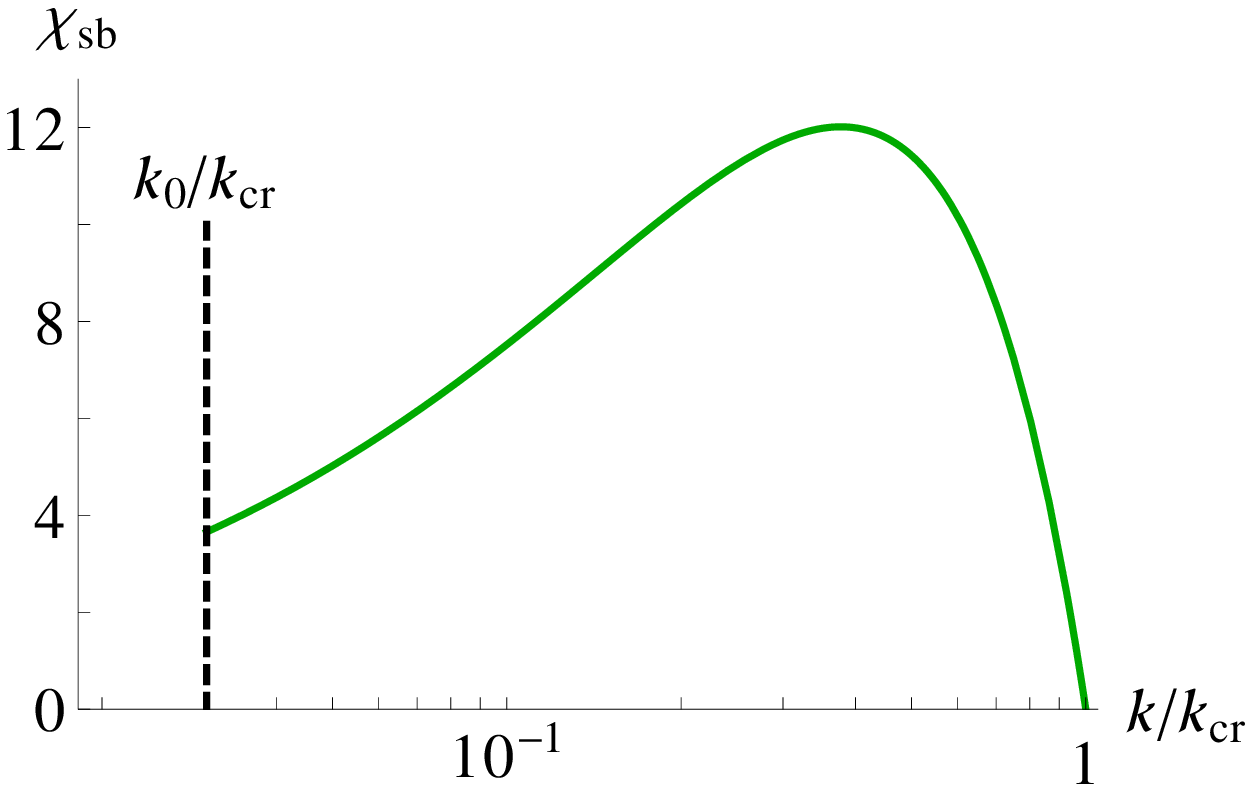}
\includegraphics[width=0.325 \linewidth]{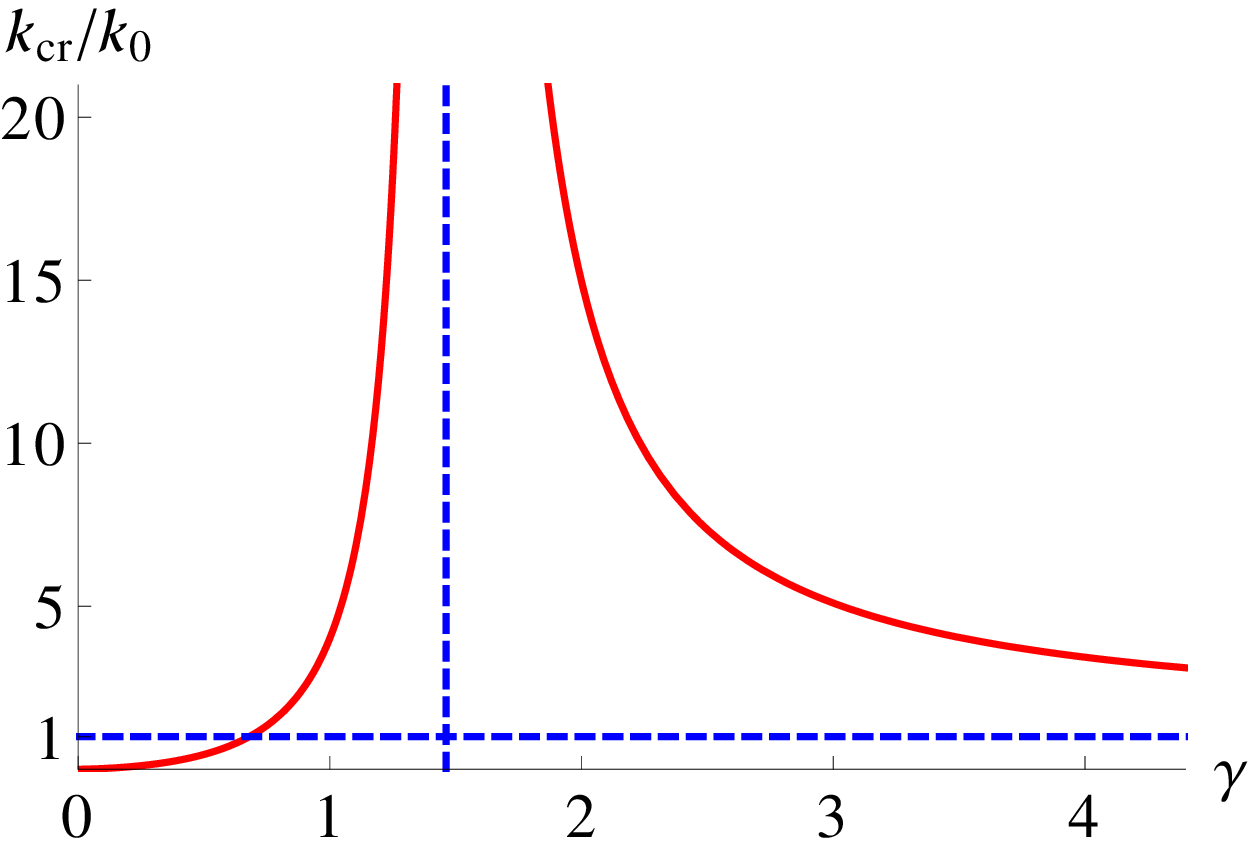}
\caption{\label{f:2} \emph{Left/Middle panels}: Mismatch functions~$\chi\sb{sp}$ (resp.~$\chi\sb{sb}$) for the supercritical trial solution (resp. subcritical) defined in the text.  \emph{Right panel}:  Evolution of the position~$k_{\mbox{\tiny cr}}$ as a function of the strength of the dissipation parameter. The vertical dashed line corresponds to~$\Gamma \equiv \Gamma_{\mbox{\tiny cr}}$ and the horizontal dashed line represents~$k_{\mbox{\tiny cr}} \equiv k_0$.}
\label{FigError}
\end{center}
\end{figure*}

\vskip 0.2cm
 \noindent  {\bf \small \emph{B. Simplified solution for the energy spectra} }
  \vskip 0.2cm

In this Section we overview some simplified solutions for the energy spectra that will serve as building blocks for the general solution.  An exact solution of Eqs.~\eqref{contHydro} with the differential closure \eqref{LN} was found in~Ref.~[\onlinecite{Nazar-Leith}] in the   case~$\Gamma = 0$.  We rewrite it as $E(\varepsilon|k)=  E\Sb{K41}(\varepsilon|k) \, T\sb{eq}(k)$ with
\begin{equation}\label{eq}
   T\sb{eq}(k) \equiv  \big[1+ \big(k/k\sb {eq}\big)^{11/2} \big]^{2/3}\ .
\end{equation}
\end{subequations} 
One can see that $E(\varepsilon|k)$ has a thermodynamic tail $E(k)\propto k^2$ for $k\gg k\sb{eq}$, as required.  A preliminary analysis into how to account for the mutual friction term in the balance equation Eq.~\eqref{contHydro} was done in Ref.~[\onlinecite{LNV}] by using the algebraical closure~\eqref{closureA} and three regimes were identified:

\textbullet~``Critical" scale-invariant solution
\begin{subequations}\label{Gamma}
 \begin{equation}\label{cs}E\sb{cr}(k)= \frac{16}{25}\,  \frac{\Gamma^2}{k^3}
 =  \Big( \frac{4 \gamma}{5}\Big )^2
\frac {\varepsilon^{2/3}_0 k_0^{4/3}} {k^3}\ .
 \end{equation}
  At the special value~$\gamma=\gamma\sb{cr}= \frac{5^{2/3}}2\approx 1.46$ it is realized for any $k$.  One sees that $\gamma\sb {cr}$ is of the order of unity and that the corresponding dimensional~$\Gamma\sb{cr}$ is about the turnover frequency of the outer-scale eddies.

\textbullet~For $\gamma>\gamma\sb{cr}$ solutions become  ``sub-critical" $E\sb{sb}(k)<E\sb{cr}(k)$  and  vanish at some finite~$k=k\sb{cr}$:
\begin{equation}\label{Ba}
E\sb {sb}  (k) \simeq  E\sb{cr}(k)\,T_-(k), \quad
T_-(k)\equiv \Big (1-  \frac{k^{2/3}}{k\sb {cr}^{2/3}} \Big )^2.
\end{equation}

 \textbullet~When $\gamma<\gamma\sb{cr}$ the energy spectra become ``super-critical" with $E\sb{sb}(k)>E\sb{cr}(k)$. For $k\ll  k\sb{cr}$, these spectra are close to the critical one and for $k\gg  k\sb{cr}$ they develop a K41 tail parametrized by~$\varepsilon_\infty< \varepsilon_0$:
 \begin{equation}\label{Ba}
E\sb{sp} (k) \simeq  E\Sb{K41}(\varepsilon_\infty|k)\,T_+(k), \
 T_+(k)\equiv \Big (1+ \frac{k^{2/3}\sb {cr}}{k^{2/3}} \Big )^2.
\end{equation}

\noindent When~$\gamma\to \gamma\sb{cr}$,  $\varepsilon_\infty\to 0$ and $k\sb{cr}\to \infty$:
\begin{equation}\label{cr}
\varepsilon_\infty=\varepsilon_0(\gamma-\gamma\sb{cr})^2\,, \quad k\sb{cr}= k_0|\gamma -\gamma\sb{cr}|^{-2/3}\ .
\end{equation}\end{subequations}

These simple solutions allow us to construct physically motivated approximate analytical solutions to~Eqs.~(\ref{contHydro}, \ref{LN}).  The accuracy of these solutions can be improved if necessary by introducing a correcting polynomial expansion to our trial energy spectra.  However, for the purpose of the present discussion it is sufficient to present only the most simple versions of the energy spectra which are determined by physical considerations only.

Notice that with the differential closure~\eqref{LN} the energy balance Eq.~\eqref{contHydro} has a scale invariant critical solution~\eqref{cs} at precisely the same value of $\gamma\sb{cr}$. Moreover, for $\gamma \lessgtr \gamma\sb{cr}$ there are sub- and super-critical solutions,  $E\sb{sb}(k)< E\sb{cr}(k)$ and  $E\sb{sp}(k)> E\sb{cr}(k)$ that we will discuss below.
\vskip 0.2cm
 \noindent  {\bf \small \emph{C. Supercritical energy spectra }}
  \vskip 0.2cm

Detailed analysis of the differential energy balance Eqs,~(\ref{contHydro}, \ref{LN}) leads us to the following form of approximate super-critical energy spectrum for $\Gamma < \Gamma\sb{cr}$:
\begin{eqnarray}\nonumber
E\sb{sp}(k)&=& E\Sb{K41}(\varepsilon|k) \Big \{\sqrt{T_+^2(k)+ T\sb{eq}^2(k)-1} -T\sb{sp}(k)\Big \}\,, \\   \label{full-sp}
 T\sb{sp}(k)&\equiv& \frac8{15} \Big( \frac{11}3\Big)^{1/3}\frac{\Gamma}{(\varepsilon\, k\sb {eq}^2)^{1/3}}\, \Big(\frac{k}{k\sb{eq}} \Big )^{7/6} \ .
\end{eqnarray}
This solution is plotted in Fig.~\ref{f:1}, middle panel. The solution~\eqref{full-sp}
is constructed from the building blocks $E\Sb{K41}(\varepsilon|k)$, $ T\sb{eq} (k)$ and $T_+ (k)$ discussed above and given explicitly by Eqs.~\eqref{K41}, \eqref{eq} and \eqref{Ba} respectively.   This solution~\eqref{full-sp} has a simple physical interpretation, which will  be clarified by spelling out the regions of k which possess different
scaling behavior; these regions are separated by $k\sb{cr}$ and $k\sb{eq}$:

 \textbullet~For $k \ll k\sb{cr}$, $T\sb{eq}\approx 1$, $T_+\approx (k\sb{cr}/k)^{4/3}\gg 1$, $T\sb{sp} \ll 1$:
 \begin{eqnarray}\label{small-k-A}
 E\sb{sp}(k)\approx C\Sb K \frac{\varepsilon^{2/3} k\sb{cr}^{4/3}}{k^3}, \quad  \varepsilon(k) \approx  \frac{15\, \varepsilon}{11 }  \Big (\frac{k\sb {cr}}{k}\Big )^2 \,,
 \end{eqnarray}
 i.e. solution has the form close to critical.

 \textbullet~For  $k\sb{cr}\ll k \ll k\sb{eq}$: $T_+(k)\approx T\sb{eq}(k)\approx 1$, $T\sb{sp}(k)\ll 1$. Thus in this region one observes the K41 spectrum $ E\Sb{K41}(\varepsilon|k)$ with the energy flux $\varepsilon$ that is related to  $k\sb{cr}$ and $\Gamma$ by the   following global constraint
\begin{subequations}\label{small-k} \begin{equation}\label{gc}
\varepsilon  - \varepsilon_0 = - \Gamma \int_{k_0}^{k_{\mbox{\tiny eq}}} E\sb{sp}(k)\,  \mbox{d}k\ .
\end{equation}
 Two other relations between these parameters are found from the boundary conditions for $E\sb{sp}(k)$ and $\varepsilon(k)$ at $k=k_0$. For example from  $\varepsilon(k_0)=\varepsilon_0$ one finds:
\begin{equation}\label{bc1}
\varepsilon_0=  \frac{15}{11} \, \Big ( \frac{k\sb{cr}}{k_0}\Big )^2\varepsilon\ .
\end{equation}\end{subequations}

 Notice that square root in Eq.~\eqref{full-sp} is constructed such that for $k\gg k\sb{cr}$  $\sqrt{T_+^2+T^2\sb{eq}-1}\approx T\sb{eq}$, thus ensuring that $E\sb{sp}(k)$ is close to the exact solution of Eqs.~(\ref{contHydro}, \ref{LN}) at $\Gamma=0$, that predict the bottleneck energy accumulation for $k>k\sb{eq}$. This wavevector should be fixed by the boundary conditions at smallest possible scale in the problem, inter-vortex distance  $\ell$. This condition is outside of the scope of the present description and currently we should consider $k\sb{eq}$ as an external parameter.
 \par Finally, the~$T\sb{sp}(k)$ term was obtained by performing an asymptotic analysis of the correction to the solution \eqref{full-sp} when~$\Gamma \neq 0$ and~$k  \gg k_{\mbox{\tiny eq}}$. \vskip 0.2cm

The accuracy of the trial solution~$E\sb{sp}(k)$ can be quantified by introducing a mismatch function ~$\chi\sb{sp}(k) = 100 \times [\frac{d\varepsilon(k)}{dk}\big/ \Gamma E\sb{sp}(k) -1]$ which measures the percentage of error of~$E\sb{sp}(k)$ when injected into~Eq.~\eqref{contHydro}.  As we can see in Fig.~\ref{f:2} (left panel), the disagreement never exceeds~15\% except in the vicinity of~$k\sb{eq}$ which has been left as an outside parameter here.  Having in mind the approximate character of the differential approach we conclude that our analytical form~\eqref{full-sp} for~$E\sb{sp}(k)$ is a good approximation to the exact solution.   Recall also, that if needed, our solution can be improved by supplementing it with appropriate polynomial corrections in the way demonstrated in Ref.~[\onlinecite{KW-T>0}].

\vskip 0.2cm
 \noindent  {\bf \small \emph{B. Subcritical energy spectra} }
  \vskip 0.2cm

  As~$\gamma \rightarrow \gamma\sb{cr}$, the supercritical spectrum smoothly connects with the critical solution~$E\sb{cr}(k)$ and indeed~$k\sb{cr} \rightarrow +\infty$ as can be seen in~Fig.\ref{f:1}, middle panel.   Eventually, we reach a point where $\gamma > \gamma\sb{cr}$ and the dissipation is so strong that the differential approximation predicts the resurgence of a finite wavevector $k\sb{cr}$ such that the energy spectrum~$E\sb{sb}(k\sb{cr})=0$ as well as the energy flux $\varepsilon(k\sb{sb})=0$ vanish. In this case, we propose the following form of approximate sub-critical energy spectrum
\begin{eqnarray}\label{solsB}
E\sb {sb}(k) =  E\sb{cr}(k)\, T_-(k) \, T \sb {sb}(k)\,,\   T\sb {sb}(k) \equiv  \Big[1-   \frac{k^{3/4}}{k\sb{cr}^{3/4}}\Big ]^2  \ .
\end{eqnarray}
This solution is plotted in Fig.~\ref{f:1}, middle panel. In addition to the building blocks already discussed above, this expression contains an extra correction $T\sb {sb}(k)$, that fixes its asymptotical behavior   close to the terminal wavevector~$k\sb{cr}$.  Conservation of energy, in the form of the global constraint
\begin{equation}
\Gamma \int_{k_0}^{k_{\mbox{\tiny cr}}} E_B(k) \mbox{d}k = \varepsilon_0,
\end{equation}
selects the value of $k\sb{cr}$, whose evolution as a function of $\Gamma$ can be seen in~Fig.\ref{f:2}, middle panel.  When~$\Gamma \rightarrow \Gamma_{\mbox{\tiny cr}}$ the position of the terminal wavevector is pushed to~$k_{\mbox{\tiny cr}} \rightarrow +\infty$ which is consistent with a smooth connection with the critical solution~$E_{\mbox{\tiny cr}}(k)$.  On the other hand, when the dissipation is very strong~$\Gamma \gg \Gamma_{\mbox{\tiny cr}}$, the terminal wavevector rapidly decreases and approaches~$k_{\mbox{\tiny cr}} \rightarrow k_0$ asymptotically as it should.

The accuracy of the trial solution~$E\sb{sb}(k)$ can be assessed by defining a mismatch function~$\chi\sb{sb}$ in complete analogy with that defined in the previous paragraph.  We can see in~Fig.\ref{f:2} (middle panel) that the error always stays below~12\%.  This observation leads us to the conclusion that~$E\sb{sb}(k)$ is a good analytical approximation to the exact solution which is more than satisfactory for our purposes.

\vskip 0.4cm

\noindent {\bf III. Discussion and conclusive remarks}

\vskip 0.2cm

 \noindent  {\bf \small \emph{A.  Numerical vs. analytical  energy spectra} }
  \vskip 0.2cm

The main results of our numerical simulations in the framework of the Sabra-shell model are presented in Fig.~\ref{f:1}.  In order to continue our discussion, notice that in the left panel one sees a horizontal line for $\gamma=0$ in the plots of $k^{5/3}E(k)$ vs. $k$.  Clearly, a K41 energy spectrum $E(k)\propto \varepsilon_0^{2/3}k^{-5/3}$ and $k$-independent energy flux $\varepsilon(k)$ which is equal to the energy influx $\varepsilon_0$ (shown in right panel) is expected in this case.  Intermittency corrections are practically invisible.

For small values of mutual friction parameter $\gamma$, we see that the energy flux first decays in the region of small $k$ before approaching $\varepsilon_\infty$ whose value is analytically estimated by Eq.~\eqref{cr}.  This behavior can be understood quite straightforwardly.  Indeed, while the mutual damping frequency $\Gamma$ is $k$-independent the K41 eddy-lifetime frequency increases as $\omega(k)\simeq \varepsilon^{1/3}k^{2/3}$.  Therefore it is always the mutual damping that must dominate the energy flux over scales in the region of small~$k$.  Such a damping naturally results in a decay of $\varepsilon(k)$.  On the other hand, in the region of large $k$ it is the characteristic frequency of energy transfer over scales, $\omega(k)$, which dominates over $\Gamma$ and therefore the mutual friction can be neglected.  In this region $\varepsilon(k)\to \varepsilon_\infty$ and the K41 scaling $E(k)\propto  \varepsilon_\infty^{2/3}$ must be recovered as is indeed confirmed both by the numerical and analytical plots of $E(k)$ in the left and middle panels.

Moreover, physical intuition [supported by Eq.~\eqref{cr}] tells us that for some (critical) value of $\gamma$ the asymptotical value $\lim_{k\to\infty}\equiv\varepsilon_\infty$ should vanish. In this \emph{critical regime} the mutual damping frequency $\Gamma$ is balanced by the cascade frequency $\omega(k)$ which can be estimated as $k \sqrt {k E(k)}$.  This immediately suggests that $E\sb{cr}(k)\simeq \Gamma^2/k^3$ which is corresponding precisely to our numerical observations (dashed red line in left panel) as well as our analytical predictions, Eq.~\eqref{cs}.

Finally, we conclude by mentioning that the numerical value of~$k\sb{cr}$ is expected to be very sensitive to~$\gamma$ in the vicinity of~$\gamma\sb{cr}$ as hinted at by~Eq.\eqref{cr}.  Indeed, the complete evolution of~$k\sb{cr}$ in both super- and subcritical regimes is presented in~Fig.\ref{f:2} (right panel).

\vskip 0.2cm

\noindent  {\bf \small \emph{B.  Mutual friction parameters $\bm \gamma$ vs. $\bm \alpha$} }

\vskip 0.2cm

The transition from subcritical to supercritical energy spectra is controlled by the dimensionles parameter $\gamma$ instead of the more experimentally accessible mutual friction parameter $\alpha$.  Both can be related to each other by Eqs.~\eqref{self-con} and \eqref{gamma}. Notice however that unlike $\gamma\sb{cr}\simeq 1$ which is universal, the critical value $\alpha\sb{cr}$ does depend on experimental parameters such as the energy injection rate at the outer scale $l$ and on the extent of the ``classical" inertial interval $l/\ell$.  It is nevertheless interesting to estimate $\alpha\sb{cr}$ on the basis of typical experimental parameters. Substituting the critical solution~\eqref{cs} in Eq.~\eqref{self-con} one gets

\begin{equation}\label{est2}
  \alpha\sb{cr}\simeq \frac{5}{4 \sqrt{2 \ln (k_0\ell)}}\simeq \frac{0.9}{\sqrt{\ln (k_0\ell)}} \ .
  \end{equation}
With $k_0\ell\simeq 10^3$~[\onlinecite{golovExp}] this gives $\alpha\sb{cr}\simeq 0.34$. For $\alpha\ll \alpha\sb{cr}$ one can neglect the effect of the mutual friction and directly substitute the K41 spectrum~\eqref{K41} into Eq.~\eqref{self-con}.  Under the assumption that this spectrum is valid up to $k\simeq 1/\ell $, one finds a remarkably simple relation for small~$\alpha$:
  \begin{equation}\label{est3}
  \gamma\simeq 2.25 \, \alpha\ .
  \end{equation}
One should be mindful that due to the possible bottleneck energy accumulation~\cite{blendingFunction}, which depends on the ratio of $\ell$ to vortex-core radius, the numerical coefficient in Eq.~\eqref{est3} can increase.  Notwithstanding these reservations, we believe that this equation should be useful, even if only as a rough estimate.

\vskip 0.2cm
 \noindent  {\bf \small \emph{C.  The road ahead} }
  \vskip 0.2cm
  To finish the discussion we should underline again the approximations that will have to be relaxed in future work.\\ \textbullet~The first one is the neglect of the turbulent motion of the normal component.
    We can study this approximation numerically, using a properly generalized~\cite{WB2011} Sabra-shell model~\cite{Sabra}, and analytically, within properly formulated differential approximation for the normal and superfluid energy fluxes.

  \textbullet~Secondly, using the two-fluid results we will generalize the model of Ref.~\cite{blendingFunction} of gradual eddy-wave crossover to the case of non-zero temperatures, to study the effect of temperature suppression of the bottleneck energy accumulation  near intervortex scales.

  \textbullet~Lastly we did not consider stability; the analysis of the stability of one- and two-fluid energy spectra against thermal perturbations should be accomplished,  having in mind the possible  self-sustained oscillations around, or in the vicinity of stationary solutions.

\vskip 0.2cm
\noindent  {\bf \small \emph{Acknowledgements.}}  This work is supported by the EU FP7 Microkelvin program  (Project No.~228464) and by  the U.S.–Israel
Binational Science Foundation.

\end{document}